\documentclass[preprint2]{aastex631}

\usepackage{subfigure}
\usepackage{comment}

\begin{document}

\title{Height-Dependent Slow Magnetoacoustic Wave Amplitude and Energy Flux in Sunspot Atmospheres}

\author{Y. Sanjay}
\affiliation{Department of Physics, School of Advanced Engineering,\\
University of Petroleum and Energy Studies, Dehradun-248007, Uttarakhand, India}

\author{S.Krishna Prasad}
\affiliation{Aryabhatta Research Institute of Observational Sciences (ARIES), \\Manora Peak, Nainital-263001, Uttarakhand, India}

\author{P.S. Rawat}
\affiliation{Department of Physics, School of Advanced Engineering,\\
University of Petroleum and Energy Studies, Dehradun-248007, Uttarakhand, India}

\begin{abstract}

Slow magnetoacoustic waves (SMAWs) have been considered in the past a possible candidate for chromospheric heating. This study analyzed 20 active regions observed between 2012 and 2016 to examine the amplitude and energy flux variation of SMAWs in the umbral atmosphere. Six different wavelength channels from the Atmospheric Imaging Assembly onboard the Solar Dynamics Observatory, covering regions from the photosphere to the low corona, were utilised for this purpose.  The wave amplitude estimations show a gradual increase in 3 minute oscillation amplitude, peaking between 700 $\--$ 900 $\textrm{km}$, followed by a steady decrease and further at altitudes greater than 1800 \textrm{km} it appears to increase and decrease again. The corresponding energy flux, on the other hand, displays a steady and monotonous decrease with a significant reduction in value from approximately 3.32 $\pm$ 0.50 \textrm{kW}{\,}$\textrm{m}^{-2}$ near the photosphere to about 6.47 $\pm$ 3.16 $\times$ $10^{-4}$ \textrm{W}{\,}$\textrm{m}^{-2}$ at an altitude of 2585 $\textrm{km}$. This decay may be attributed to radiative damping and shock dissipation in the lower altitudes and thermal conduction and viscosity in the higher altitudes. The missing flux is a factor of 3 – 15 lower than that required to counterbalance the chromospheric radiative losses.

\end{abstract}


\section{Introduction} \label{sec:intro}
The coronal heating problem was discovered around eighty years ago, notably by \citet{1939NW.....27..214G} and \citet{1943ZA.....22...30E}. Their identification of spectral lines such as Fe IX and Ca XIV within the solar corona highlighted the presence of anomalous temperatures in the multi-million-degree range. It is widely acknowledged that the energy fueling the corona's extreme temperatures originates from the convective gas motions occurring in the solar photosphere. The lower solar atmosphere (photosphere and chromosphere) plays an important role in transferring this energy from the solar interior to the corona. Magnetohydrodynamic (MHD) waves offer one of the key mechanisms to enable this transfer. Therefore, determining the height-dependent variation in wave energy flux is a crucial step in order to address the coronal heating problem. 

The energy loss in the solar atmosphere is primarily due to radiation, thermal conduction, and mass transport. Radiation is the primary means of energy loss in the chromosphere, while in the corona, all three mechanisms  $\--$ radiation, thermal conduction to the lower atmosphere, and mass transfer into the heliosphere via the solar wind $\--$ contribute to energy loss \citep{1977ARA&A..15..363W}. Thermal conduction and plasma flow in sunspots are restricted to follow magnetic field lines. Moreover, it is well established that the solar chromosphere needs a significantly higher energy flux than the corona (e.g., total energy losses in active region atmosphere are 10 $\textrm{$kW\ m^{-2}$}$ in the corona and 20 $\textrm{$kW\ m^{-2}$}$ in the chromosphere \citep{1977ARA&A..15..363W}) because of the larger mass of the chromosphere. However, despite this understanding, the exact mechanism(s) by which the required energy (to replenish the losses) is transported and dissipated within the solar atmosphere is not yet fully comprehended.

It is currently believed that atmospheric heating is mainly due to MHD waves \citep[AC model;][]{1947MNRAS.107..211A} and magnetic reconnection \citep[DC model;][]{1947MNRAS.107..338G}. The convective plasma flows in the photosphere interact with the magnetic fields resulting in either the excitation of MHD waves or the tangling of the magnetic fields which will eventually release the energy via reconnection \citep{1974ARA&A..12..407S}.

In a homogeneous medium with a uniform magnetic field,  three types of MHD waves exist: the Alfv\'en waves \citep{1942Natur.150..405A}, the slow and the fast magnetoacoustic waves (MAWs) \citep{1974ARA&A..12..407S}. The Alfv\'en waves are most difficult to dissipate because of their incompressive nature. In contrast, the dissipation of compressive slow MAWs is much easier.

According to \citet{1977A&A....55..239B}, slow waves cannot travel against gravity below a cutoff frequency. So, the 5 minute oscillations are the most prevalent in the photosphere of a sunspot but as one rises into the chromosphere and corona the 3 minute becomes the dominant period \citep{1987ApJ...312..457T}. In the umbral atmosphere where magnetic field lines are mostly vertical, the 3-minute slow MAWs propagate along magnetic field from the photosphere to all the way to the corona. The decreasing density in the solar atmosphere causes their amplitude to grow and they may evolve into shock waves due to steepening. The subsequent dissipation might contribute to the heating of the atmosphere \citep{2004Natur.430..536D,2008A&A...479..213B}. The
sawtooth pattern observed in the Doppler velocity profiles in chromosphere indicates that this shock development indeed takes place \citep{2014ApJ...786..137T}. According to \citet{1989ApJ...336.1089A}, the slow MAWs generated at the photosphere are inadequate to heat the solar corona because of this quick dissipation, making them a possible candidate for chromospheric heating.

\citet{1981A&A...102..147K} found that the acoustic energy fluxes of 18 $\textrm{$W\ m^{-2}$}$ and 49 $\textrm{$W\ m^{-2}$}$ estimated from the root-mean-squared (RMS) velocities of the $Na_1 \ D_1$ and $D_2$ lines respectively, were insufficient to balance chromospheric radiation losses. The authors considered a height separation of 110 $\textrm{km}$ between the formation of two lines and determined the group velocity using the time lag obtained between these two layers. However, they explicitly note that the umbral density $10^{-11} \textrm{$g\ cm^{-3}$}$ used in the calculations, was not well known at that time, which therefore, adds uncertainty to their conclusions. 

In a later work, data-driven MHD simulations revealed that the average acoustic energy reaching the chromosphere was about 3 to 9 times lower than the energy required for sustaining chromospheric heating \citep{2011ApJ...735...65F}. A similar conclusion was reached by \citet{2017ApJ...836...18C}, who applied the lambda-meter method to determine velocity amplitudes, and reported that the energy flux for 3 minute oscillations at a height of 38 $\textrm{km}$ was only 1.8 $\textrm{$kW\ m^{-2}$}$, which is already insufficient to maintain the umbral chromosphere.

A more recent study using Stokes inversion with the Milne-Eddington approximation estimated that the difference in energy flux between the photosphere and the lower transition region in the 6 $\--$ 10 $\textrm{mHz}$ band was around 20 $\textrm{$kW\ m^{-2}$}$ \citep{2016ApJ...831...24K}. While this energy flux is adequate to maintain the chromosphere, their estimated densities were higher than those of the standard empirical umbral model by \citet{1986ApJ...306..284M}. The authors noted that if the model densities were applied, the calculated energy flux would fall below the chromospheric heating requirement. 

Accurate energy flux determination is critical on three parameters $\--$ mass density, velocity amplitude, and wave propagation speed. \citet{2017ApJ...847....5K} estimated the height-dependent relative intensity amplitudes of 3-minute waves using multi-wavelength data from the Dunn Solar Telescope and the Interface Region Imaging Spectrograph \citep[IRIS;][]{2014SoPh..289.2733D}. In order to obtain the corresponding velocity amplitudes, the authors scaled the intensity amplitude estimated from the Rapid Oscillations in the Solar Atmosphere \citep[ROSA;][]{2010SoPh..261..363J} blue continuum observations to the RMS velocity (40 $\textrm{$m\ s^{-1}$}$) reported by \citet{1985ApJ...294..682L}, measured at $\approx$ 40 $\textrm{km}$ above the 
photosphere. Furthermore, considering the temperature and density values from the 
umbral 'M' model of \citet{1986ApJ...306..284M}, they find a significant 
damping of energy flux from about 13 $\textrm{$kW\ m^{-2}$}$ at the photosphere to about 0.7 $\textrm{$W\ m^{-2}$}$ at a height of $\approx$ 1500 $\textrm{km}$.

Recently, \citet{2024MNRAS.533.1166R} determined the amplitude of velocity oscillations at umbral fan loop locations to be about $\approx$ 33  $\textrm{$m\ s^{-1}$}$ near the photosphere. The authors utilise the 3 minute filtered Dopplergram data from the Helioseismic Magnetic Imager \citep[HMI;][]{2012SoPh..275..207S,2012SoPh..275..229S} onboard the Solar Dynamic Observatory \citep[SDO;][]{2012SoPh..275....3P} for this purpose. In addition, using IRIS spectroscopic observations from $\textrm{Mg II}$ 2796 \r{A} line, the authors determined the velocity amplitude in chromosphere is about 1 $\textrm{$km\ s^{-1}$}$. This value is notably much higher than the estimates provided by \citet{2017ApJ...847....5K}. The velocity amplitudes at other atmospheric heights, corresponding to the Atmospheric Imaging Assembly  \citep[AIA;][]{2012SoPh..275...17L} data from 304 and 171 \r{A} channels, were determined by assuming a linear scaling between the intensity and velocity amplitudes \citep{2009A&A...503L..25W}. The necessary density and temperature values are taken from the sunspot atmosphere model of \citet{1999ApJ...518..480F}. The resultant energy flux values were found to be similar to that of \citet{2017ApJ...847....5K} i.e., about 10 $\textrm{$kW\ m^{-2}$}$ at photosphere and about  2 $\textrm{$W\ m^{-2}$}$ near the low corona.

Notably, the previous studies mainly rely on various atmospheric models for determining the umbral temperature and density values which are not necessarily applicable for all active regions. Indeed, these parameters are expected to vary across different active regions and heights but are very challenging to measure. In this study, we aim to estimate these parameters as independently as possible and calculate the height-dependent wave energy flux variations for 20 different active regions. 

In Section{\,}\ref{sec:data}, we describe the details of the datasets employed, in Section{\,}\ref{sec:results} we describe the analysis methods adopted and the results obtained and finally, in Section{\,}\ref{sec:conclusion} we present our conclusions.

\section{Data}\label{sec:data}
The dataset used in this study is a subset of the data from \citet{2018ApJ...868..149K} and is identical to that employed in \citet{2024ApJ...975..236S}. It includes level 1.5 subfield data (180\arcsec $\times$ 180\arcsec) of 1 hr duration from SDO/AIA and SDO/HMI, spanning 20 different active regions (AR) observed between 2012 and 2016. The specific wavelength channels used are the HMI Continuum, AIA 1600, 1700, 304, 131, and 171 \r{A}. The individual AR datasets comprise 80 images for the HMI data, 149 images for AIA 1700 \r{A}, 150 images for AIA 1600 \r{A}, and 300 images each for the remaining EUV channels. The respective time cadences are 45 s for the HMI data, 24 s for the AIA 1600 and 1700 data, and 12 s for the other three EUV channels. Additionally, we use HMI Dopplergram data for wave amplitude scaling and HMI magnetogram data for context purposes. The pixel scale is the same across all the data which is about 0.6$\arcsec$. The start time and the corresponding NOAA AR number for each dataset are listed in Table \ref{tab:density}. Since we are interested in oscillations in the umbral region, it must be noted that about eight of the sunspots identified display light bridges (marked by asterisks in the table). However, we restricted our analysis to the umbral region after carefully excluding the light bridge locations. Additionally, no significant flare-like activity was found in any of the ARs during the selected period.

\section{Analysis and results}\label{sec:results}
\begin{figure*}
    \centering
    \includegraphics[width = \linewidth]{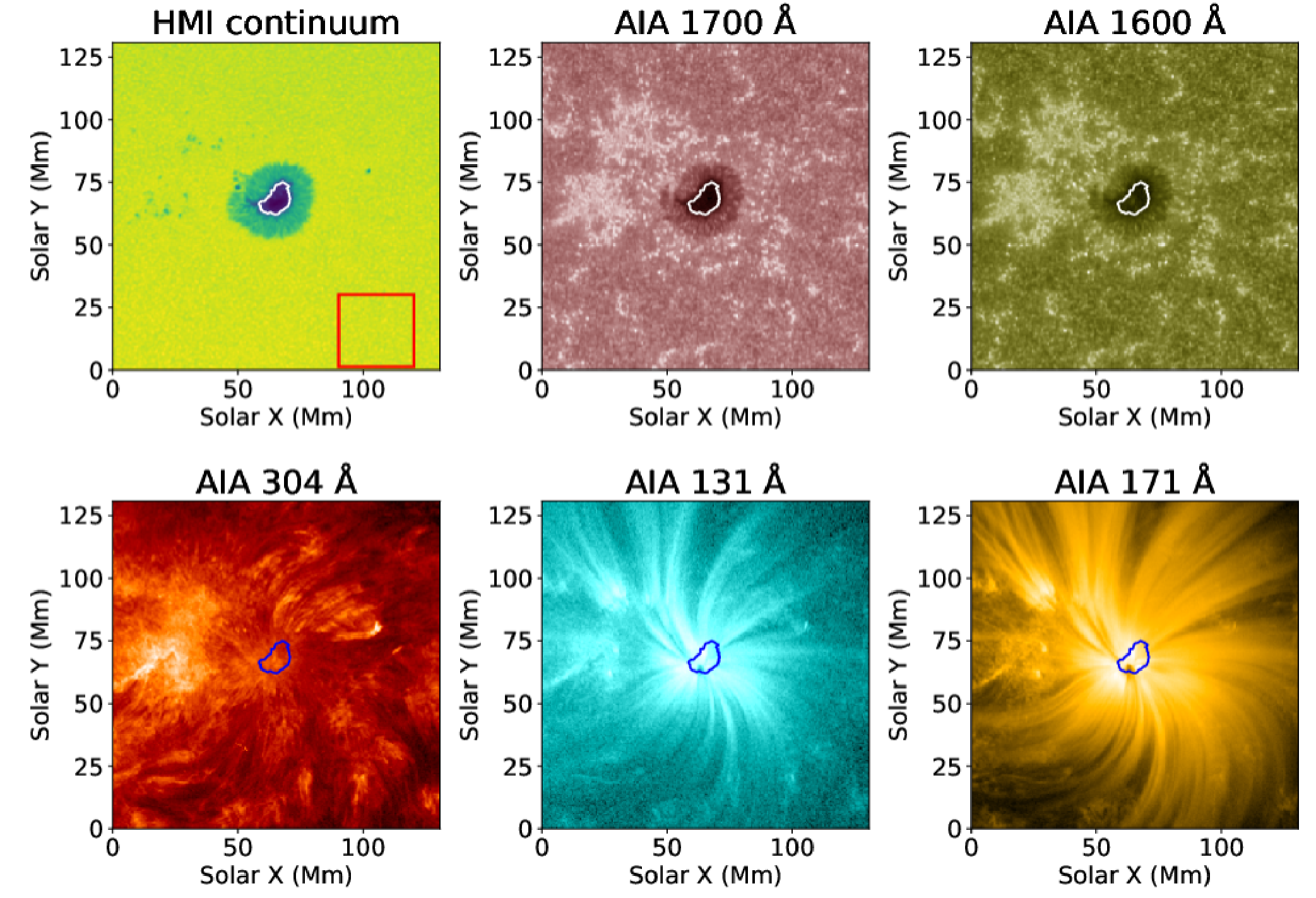} 
    \caption{
Sample images of a sunspot within AR 12005 captured on 2014 March 18, across various wavelength channels of SDO as shown. The first image (HMI continuum) represents the time-averaged intensity over a one-hour duration, while the rest depict the first frame within their respective image sequences. The red rectangular box in the HMI continuum denotes the selected quiet region for calculating the median intensity. The white/blue solid curves delineate the boundary between the umbra and penumbra.}
    \label{fig:umbra}
\end{figure*}
Our main objective of this study is to obtain the wave amplitude and energy flux of Slow Magneto Acoustic Waves (SMAWs) as a function of height in the solar atmosphere. In order to achieve this, within each of the AR datasets, we identify a prominent sunspot and then analyze the oscillations within the corresponding umbral region using multi-wavelength imaging data from SDO/AIA and SDO/HMI. The reason for limiting our analysis to the umbral region is the near vertical field expected in the umbra which ensures that we are following the same structure across different heights in the atmosphere.

The first step, therefore, is to select the umbral region after excluding light bridge locations, if any, present. We consider HMI continuum data and employ a simple intensity thresholding technique for this purpose \citep{2013ApJ...779..168J}. Typically, the time-averaged median intensity over a quiet region away from the sunspot is computed, and a fixed fraction of this value is applied as a threshold until a reasonable match in the umbra-penumbra boundary is obtained through visual inspection. Across the 20 active regions, this fraction was found to lie between 20\%$\--$ 55\%. The obtained boundary between the umbra and penumbra of a sunspot within AR 12005 is shown by white/blue contours in Fig. \ref{fig:umbra}. The quiet region selected away from the sunspot is shown by a red box over HMI continuum image in this figure.

In the next step, we seek to obtain the average wave amplitude and energy flux within the umbral region. While the amplitude of a wave is an independent parameter, the energy flux additionally requires knowledge of local density and wave propagation speed. We describe our methodology to extract all these parameters in the following subsections.

\subsection{Wave amplitudes}
The amplitude of SMAWs is estimated across multiple heights following the methodology described in \citet{2017ApJ...847....5K}. First, the time series from each pixel location within the umbral region is subjected to Fourier analysis to obtain the corresponding power spectrum. A mean power spectrum is then constructed for each wavelength channel by simply averaging all the power spectra within the umbral region. Note that we are not looking for global oscillation modes within the umbra but rather intend to find the average behavior of the oscillations. 

\begin{figure*}
    \centering
    \includegraphics[width = \linewidth]{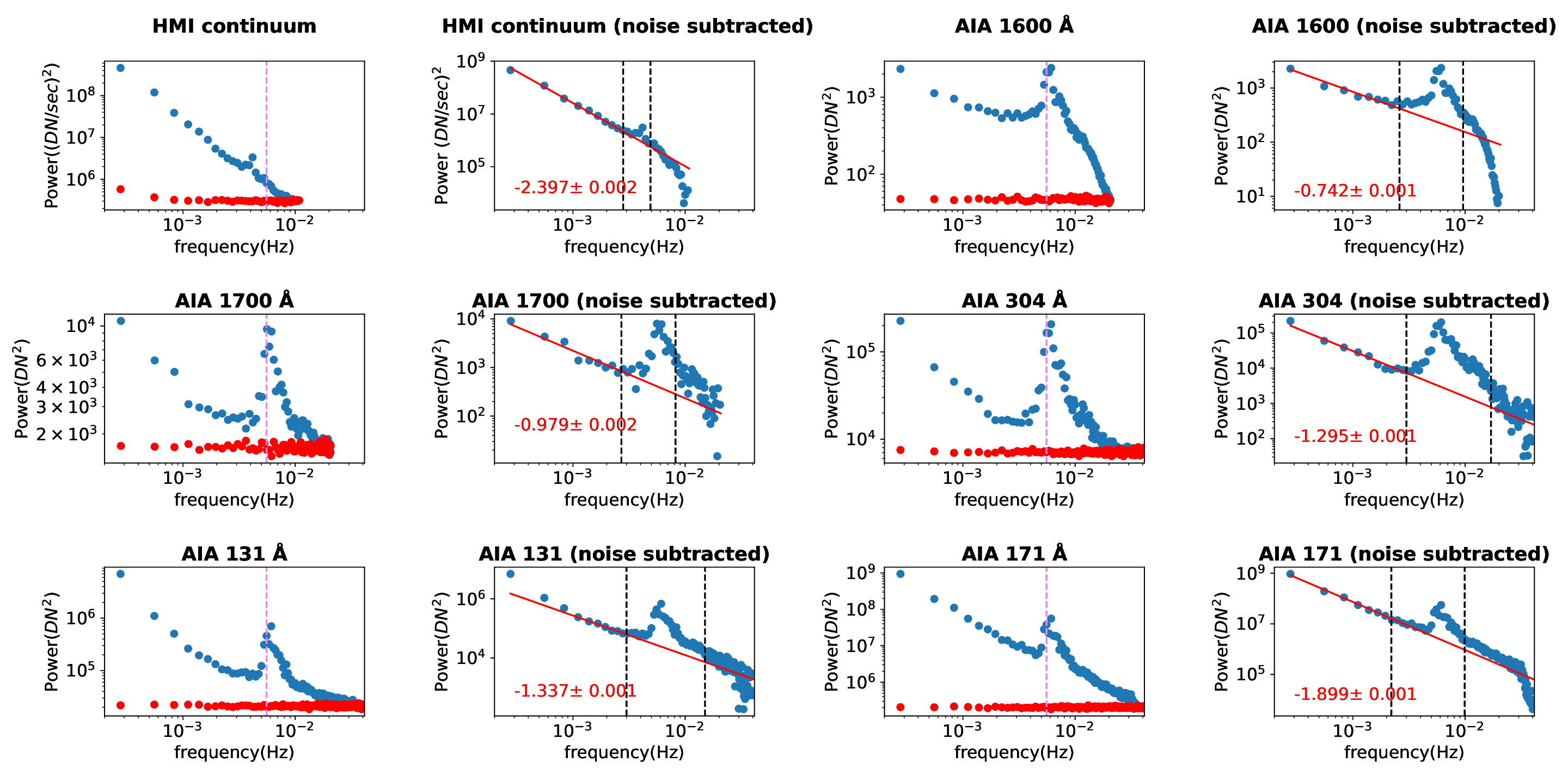}
    \caption{Average Fourier power spectra within the umbral region of a sunspot from AR 12005. The plots in the first and third columns depict the obtained spectra from 6 different SDO channels. The blue dots in these panels represent the spectra while the red dots represent the corresponding white noise estimated from observations. The vertical dashed line in violet represents the location of 5.56 mHz frequency. The plots in the second and fourth columns depict the noise-subtracted spectra in respective channels. The red solid line represents a linear fit to the data obtained after excluding the peak region identified by the two vertical dashed lines in black. The obtained slopes are listed in the plot.}
    \label{fig:noise}
\end{figure*}

The mean power spectra thus obtained for all 6 wavelength channels, are shown in Fig. \ref{fig:noise} for AR 12005 (first and third columns). All these spectra, except that from the HMI continuum, show a prominent peak near 5.6 mHz (3-minute period). The HMI continuum data display a peak at a slightly lower frequency, i.e., in the 3$\--$4 mHz (5-minute period) range. Also, the peak in HMI continuum is quite marginal. This is because of the smaller oscillation amplitudes in the photosphere, which are often further attenuated by the opacity effects \citep[e.g., see review][]{2015LRSP...12....6K}. 

The power-law behavior of the spectra appears to plateau at the high-frequency end. The plateauing is mainly because of the white noise, which dominates at high frequencies. Assuming enough signal in the data, this white noise can be modeled as photon noise and can be computed directly from the data. 
We estimated the white noise for each pixel by generating random light curves following a Poisson distribution, where the amplitude was set equal to the square root of the mean intensity at that pixel. We then calculated the mean Fourier power from all the artificially generated light curves to obtain the white noise level.
The computed white noise power, shown as red dots in the respective panels, is then subtracted from the power spectra to reveal the underlying trend at high frequencies. The noise-subtracted power spectra are shown in the second and fourth columns of the figure. 

The peak power appears to vary from channel to channel, but a direct comparison is not possible here as the corresponding intensities, which are used to compute these spectra, are not determined in absolute units. In order to enable such a comparison, each of these spectra is linearly fitted (in log-log scale) to identify a background level. The background fits are shown by red solid lines in the figure. The corresponding slopes, which represent power-law indices in the linear scale, are also listed. To obtain a better background fit, the data points surrounding the peak (between the vertical dashed lines) are excluded. Additionally, the linear fit is expected to be considerably biased towards the high-frequency data because there are far more data points at high frequencies than at low frequencies. Therefore, we applied a differential weighting with 10 \% weight for high-frequency data and 100 \% weight for low-frequency data. 

\begin{figure*}
    \centering
    \includegraphics[width = \linewidth]{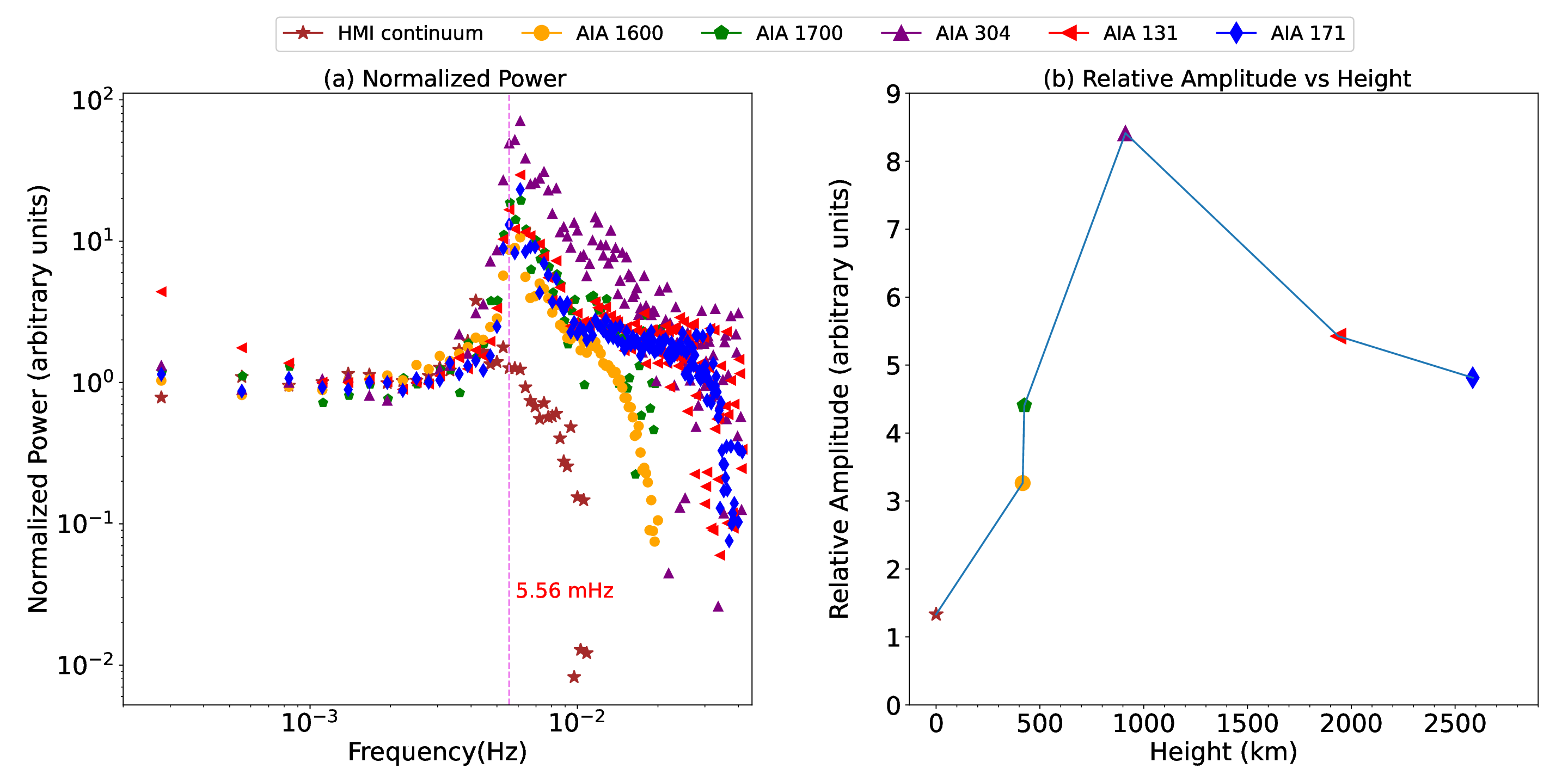}
    \caption{
    \textit{Left}: Normalized power spectra obtained for AR 12005, observed on 2014 March 18, across 6 different SDO channels. The vertical dashed line represents a frequency of 5.56 mHz. \textit{Right}: The height-dependent variation in the relative oscillation amplitude in the 3-minute band.}
    \label{fig:amplitude}
\end{figure*}

Subsequently, all the spectra are normalised with their corresponding background providing us with a relative scale. The normalised power spectra, across all 6 wavelength channels, are shown in Fig. \ref{fig:amplitude}a. Different channels are represented by unique colors and symbols in this plot. As noted previously, all the channels, barring HMI continuum, display a peak near 5.56 mHz. The square root of the power at this frequency essentially represents the amplitude of 3-minute oscillations. The variation of this amplitude as a function of height is shown in Fig. \ref{fig:amplitude}b. Note that, to enable a proper comparison and assessment of the amplitude variation, the peak power in the frequency band 4.17 - 8.33 mHz (2-4 minutes) is used to compute the amplitude for all the channels, including the HMI continuum. The formation height of individual channels has been previously determined by \citet{2024ApJ...975..236S} for all the active regions analysed in this study. Therefore, the relevant height values for AR 12005 are used in plotting the amplitudes in Fig. \ref{fig:amplitude}b. 
 
\begin{figure*}
    \centering
    \includegraphics[width = \linewidth]{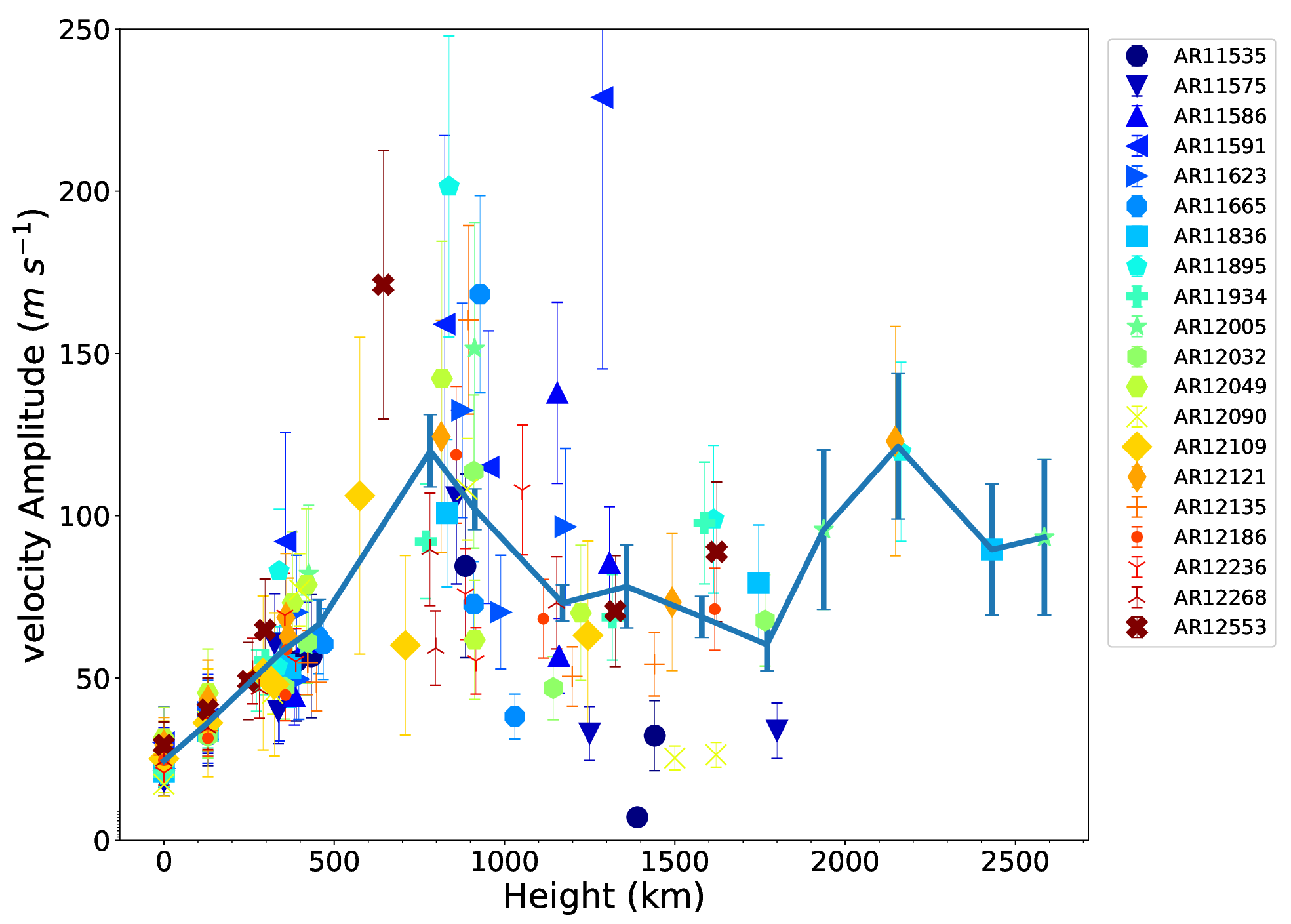}
    \caption{The height-dependent variation in the obtained velocity amplitudes for 3 minutes SMAWs. Different colors/symbols correspond to data from different active regions as listed. The corresponding formation heights are taken from \citet{2024ApJ...975..236S}. The Doppler velocity amplitudes obtained from the HMI Dopplergram data are also shown here. The solid line connects the average amplitudes and formation heights computed within each 210 km block above the HMI continuum height. The vertical bars on these data represent the corresponding standard error in the values.}
    \label{fig:ampheight}
\end{figure*}

The oscillation amplitude increases with height and reaches a peak value in the AIA 304 \r{A} channel before showing a decreasing trend at higher altitudes. 
Note that these values are on a relative scale but not yet in physical units. We need to convert them to physical space in order to use them in the energy flux calculations. To accomplish this, we use HMI Dopplergram data. 
The Doppler velocity time series at each pixel within the umbral region is filtered in the 3-minute band using the Butterworth filter, and their corresponding velocity amplitude is determined from their RMS value (peak value $= \sqrt{2}\times RMS$). 
The average velocity across the whole umbral region and the associated standard deviation are mentioned in Table \ref{tab:density} for all active regions. For AR 12005 this value is about 33.94$\pm$8.7 m s$^{-1}$. The formation height of HMI Dopplergram is $\approx$ 150 km \citep{2011SoPh..271...27F,2013SoPh..282...15C}. 
According to \citet{2006ASPC..358..193N}, the continuum near Fe{\,}\textsc{i} 6173 \r{A} (HMI continuum) forms at an altitude of $\approx$ 21 $\textrm{km}$ in the umbral atmosphere. Therefore, we use linear interpolation on the intensity amplitudes to find the amplitude value at the Dopplergram formation height, assume this to be equivalent to the velocity amplitude obtained from the Dopplergram data, and, subsequently, scale all the other amplitude values as per this relation. 
The corresponding uncertainties are estimated from error propagation using standard deviation in the Doppler velocity amplitude. 

The final height-dependent velocity amplitudes obtained for all the 20 active regions analysed in this study are presented in Fig. \ref{fig:ampheight}. These values are arranged as per the respective formation heights determined by \citet{2024ApJ...975..236S}. The velocity amplitudes obtained directly from the Dopplergram data are also included in this plot. Note that individual colors/symbols in this figure represent data from a specific active region, as listed in the legend. The data appears a bit dispersed emphasizing the differences between different active regions. Nevertheless, a general pattern may be observed. In order to highlight this, we compute the average amplitude and height of all the data points within each 210 km block over the entire height range. These values are shown using a solid line in the figure. The corresponding uncertainties (shown by thick error bars) are computed from the ratio of the standard deviation to the square root of the number of data points within each block. The amplitudes within the HMI continuum data are considered as a starting point and are placed at 0 km height. The specific choice of 210 km is only selected to effectively represent the observed variation pattern across the entire height range. The resultant curve illustrates a notable trend depicting an increase in amplitude up to a height of $\approx$ 800 $\textrm{km}$, where it reaches a peak and gradually decreases at higher altitudes. The decreasing trend continues up to $\approx$ 1800 $\textrm{km}$ and, interestingly, starts to increase and decrease again at further heights. 

It may be noted that the height of peak amplitude obtained here is different from \citet{2017ApJ...847....5K} where the authors find the wave amplitude peaking at a height of 1300 km corresponding to the formation of Ca{} II K line core. This discrepancy is likely because we use AIA 304 \r{A} channel here to represent the chromosphere instead of Ca{} II K core. It is possible that Ca II K core is forming higher than AIA 304 \r{A} channel. Still, we would also like to point out that the height of Ca {} II K core in \citet{2017ApJ...847....5K} is largely determined from the previous literature. In contrast, \citet{2024ApJ...975..236S} determined the formation height of AIA 304 \r{A} using a comprehensive phase difference analysis for the same active regions for which we measured the wave amplitudes. 
Furthermore, our analyses include 20 different active regions, providing a broader statistical sample and minimizing the influence of individual cases, leading to more generalizable results. The formation height of AIA 304 \r{A} in our dataset spans approximately 575 $\pm$ 40 km to 1155 $\pm$ 48 km \citep{2024ApJ...975..236S}, depending on the specific active region, contributing to differences in the observed peak locations.

Additionally, it is important to acknowledge the complexity of  AIA 304 \r{A} passband. It captures emission primarily from cool, dense plasma (He{} II $\approx 10^5$ K), but also has additional contributions from hotter lines like O{} V ($\approx$ 0.3 MK) and Si XI ( $\approx$ 2 MK). However, the hotter components are primarily applicable for off-limb regions \citep{2024SoPh..299...94A}. Indeed, \citet{2000SoPh..195...45T} used the data from Coronal Diagnostic Spectrometer (CDS) onboard the Solar and Heliospheric Observatory (SOHO) to show that Si{} XI emission is an order of magnitude weaker than He{} II on-disk, but it becomes comparable for off-limb observations due to longer line-of-sight (LOS) integration paths through hotter plasma. In our study, all observations were made near the disk center. Therefore, we believe that the broad temperature response of AIA 304 \r{A} channel has a lesser impact on our results.

\subsection{Propagation speed}
	\label{sec:speed}
The propagation speed of SMAWs is generally given by the tube speed, but under low plasma-$\beta$ conditions, this value is pretty close to the local sound speed. Since we are dealing with the sunspot umbral atmospheres, low plasma-$\beta$ is a reasonable approximation. As we will see later, the plasma-$\beta$ for our active region sample is about 2.5 in the photosphere. This implies that the difference between the tube speed and the sound speed will be marginal so we approximate the local sound speed as the propagation speed of SMAWs observed here.

The local sound speed can be calculated using the formula:
 \begin{equation}
    c_s = \sqrt{\frac{\gamma \  k_B \ T}{\mu \ m_H}},
    \label{sound}
\end{equation}
here, $k_B$ = Boltzmann's constant, $T$ = Temperature in kelvin,
$\mu$$m_H$ =  mean mass of the particle with $\mu$ = 0.61 \citep{1992str..book.....M} for corona and 1.25 for photosphere. The selection of temperatures for different channels of HMI and AIA is comprehensively described in \citet{2024ApJ...975..236S} and we adopted the same values in this study. To explain briefly, the characteristic temperature of AIA EUV channels (304, 131, and 171 \r{A}) is taken as the peak of their respective temperature response curves appropriate for non-flaring conditions, the characteristic temperature for the AIA UV channels (1600, 1700 \r{A}) is considered to be same as photospheric continuum based on the dominant continuum contribution to these channels, and temperature of the HMI continuum is estimated using the observed umbral intensities assuming the emission as black body radiation. 

The obtained temperature values are 5011 K, 5011 K, 50118 K, 0.45 MK, and 0.79 MK for the AIA 1600, 1700, 304, 131, and 171 \r{A} channels respectively. The corresponding sound speeds are  7.4, 7.4, 33.6, 100.6, and 133.3 km s$^{-1}$. While these values are fixed, the HMI continuum temperature is different for different active regions as this is dependent on the observed umbral brightness. This value ranges from 4176 $\--$ 4764 K and the corresponding sound speed varies from  6.8$\--$7.2 km s$^{-1}$.

\subsection{Density}
Extracting height-dependent plasma densities is not possible from intensity observations alone. Instead, spectroscopic/spectropolarimetric observations are required. But such observations, including multiple wavelength channels, are rare, and therefore, in the past, researchers have used expected densities from various models. However, indirect methods are available to estimate at least the photospheric density from the cutoff frequency as demonstrated previously by \citet{2017ApJ...837L..11C}. Assuming a non-isothermal stratified medium with a uniform vertical magnetic field, which is applicable to the umbral region, the expression for angular cutoff frequency, $\omega_c$, in the case of finite $\beta$ plasma is as follows \citep{2006RSPTA.364..447R, 2017ApJ...837L..11C}:
\begin{equation}
   \omega _c ^2  = \frac{3\gamma  g  C^3 }{4 \lambda _ P} - C^2 \left(\frac{\gamma g}{\lambda _ P}  +  \frac{ g}{2\ \lambda _P} \right) + \frac{Cg}{ \lambda _P},
   \label{cutoff}
\end{equation}
 where $C = \frac{c_t ^2}{c_s ^2}$, $c_t$, and $c_s$ are the tube speed and sound speed, respectively, at the photosphere, $\lambda _P $ $(= \frac{c_s ^2}{\gamma g})$ is the pressure scale height, g (= 274 $\textrm{$m\ s^{-2}$}$) is the acceleration due to gravity and  $\gamma\  (= \frac{5}{3})$ is the adiabatic index. The idea is to obtain the cutoff frequency from observations, and, using the sound speed calculated from the characteristic temperature, estimate the tube speed by solving Eq. \ref{cutoff}. As the tube speed is only dependent on the Alfv\'{e}n speed and the sound speed, one could derive the Alfv\'{e}n speed and, subsequently, utilising the magnetic field strength information from the HMI magnetogram data, the corresponding gas density can be estimated.

\citet{2017ApJ...837L..11C} derived the cutoff frequency from a weighted frequency (equivalent to the peak frequency) obtained from the power spectra, by applying somewhat arbitrary scaling to the weighted frequency, which, according to the authors, gives a meaningful solution to Eq. \ref{cutoff}. Here we follow a different approach, and derive the cutoff frequency directly from the phase difference map obtained for the HMI continuum and the AIA channel immediately close to it as per the formation height \citep[this can be 1700 or 1600 \r{A} depending on the active region; see][for details]{2024ApJ...975..236S}.

\begin{figure*}
    \centering
   \includegraphics[width = \linewidth]{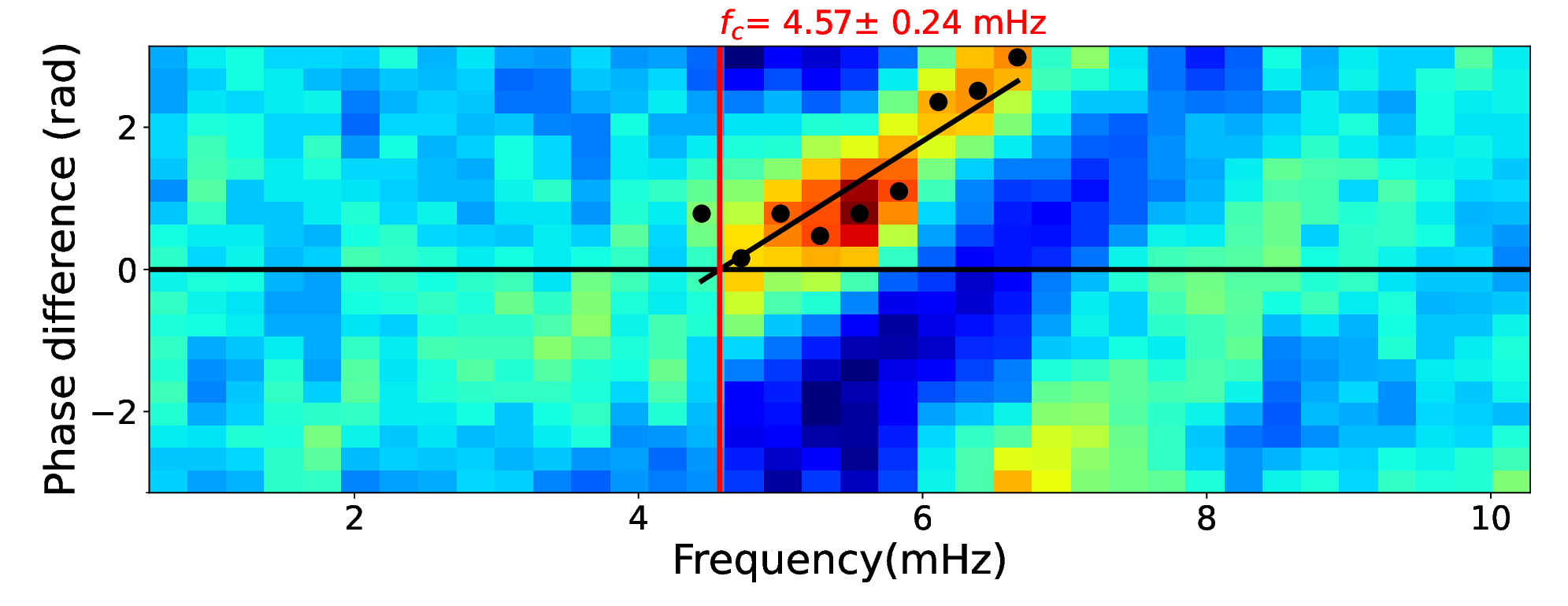}
    \caption{A map of phase difference spectra obtained across the whole umbral region within a sunspot from AR12109 observed on 2014 July 7. The phase differences are determined from a cross-power spectrum between the HMI continuum and AIA 1600 \r{A} channels. The color scale represents the occurrence rate of a particular phase for each frequency, with red and blue colors depicting high and low occurrence rates, respectively. The black dots represent the local maxima identified and the black solid line is a linear fit to these data (see text for details). The cutoff frequency obtained from the intersection of the fitted line with the zero phase line (horizontal black line) is marked by a vertical red line and the corresponding frequency value is listed.}
    \label{fig:phase}
\end{figure*}

A sample phase difference map obtained for the largest sunspot in our dataset, within AR 12109, is shown in Fig. \ref{fig:phase}. In this case, the cross-power spectra between the HMI continuum and the AIA 1600 \r{A} channel are used to determine the phase difference values. The AIA data were resampled to match the cadence of the HMI continuum. Each column in the map represents a histogram of the phase difference values obtained at that particular frequency, across all the umbral pixels.
Therefore, the color scale simply shows which value is more prevalent than the other, with red denoting values that occur more frequently and blue denoting values that occur less frequently. The histograms at each frequency are built considering 20 bins i.e., $\approx$ 0.3 rad per bin. An increasing phase in this map indicates propagating waves while a zero phase represents standing or evanescent waves \citep{2006ApJ...640.1153C, 2010ApJ...722..131F}. Although the spectrum appears somewhat noisy at the beginning, the propagating phase is still discernible at frequencies beyond 4.2 mHz. The next task is to determine the boundary at which the standing phase transitions to a propagating phase, which gives us the cutoff frequency.

In order to achieve this, starting from 4.2 mHz, at each frequency bin, we identify the positive phase at which the occurrence value is peaking until we approach a value of $\pi$ radians. In the case of multiple peaks with the same occurrence value, we consider an average between them. The locations of these peaks are shown by black dots in Fig. \ref{fig:phase}. A linear fit is then applied to these data and the frequency at which the fitted line intersects the zero phase difference line is considered as the cutoff frequency. The uncertainty in this parameter is derived from the covariance matrix obtained from the linear fitting process. The cutoff frequency ($f_c$) thus obtained for AR 12109 is 4.57$\pm$0.24 mHz. The location of this is shown by a vertical red line in the figure. The corresponding angular frequency will be 2$\pi$ times this value. The $f_c$ values obtained for all 20 active regions are listed in Table \ref{tab:density}. These values mainly lie in the 4$\--$ 5 mHz range, close to the peak observed in the HMI continuum power spectra, in agreement with previous studies \citep[e.g., see][]{2012ApJ...746..119R,2018A&A...617A..39F}.

The temperature and, thereby, the local sound speed within the HMI continuum are derived as described in Section \ref{sec:speed}. Now that we have the cutoff frequency and the sound speed values, we solve Eq. \ref{cutoff} and seek a positive solution for $C$, where $C<1$. Among the three possible solutions, we find that only one is real. In some cases, the exact real solution marginally exceeds 1 which is not physical. Therefore, we consider the closest positive solution that satisfies $C<1$. From the parameter $C$, we compute the tube speed $c_t$, the Alfv\'{e}n speed $v_A$ $\left(=\frac{c_t\ c_s}{\sqrt{c_s^2 - c_t^2}}\right)$, and the plasma $\beta$ $(=\frac{2\ c_{s}^2}{\gamma\ c_{A}^2})$. Subsequently, employing the average umbral magnetic field $(\left |B_{umbra}\right |)$ obtained from the HMI magnetogram data, we compute the photospheric density $\rho$ $\left(=\frac{\left|B_{umbra}\right|^2}{\mu_0\ c_A^2}\right)$. The corresponding uncertainties in each of these parameters are estimated using standard error propagation methods. The magnetic field, density, and plasma $\beta$ values obtained for all the active regions are listed in Table \ref{tab:density}.

Additionally, in order to estimate height-dependent densities within each active region, we use the umbral model M of \citet{1986ApJ...306..284M}. Essentially we fit the model values with a cubic spline function in a logarithmic scale and then estimate the densities at all desired heights. All these values are then scaled to match the photospheric density at 21 km (HMI continuum height) we estimated from observations. The final values obtained are in the same range as those previously reported in literature \citep[e.g.,][]{2017ApJ...837L..11C, 2018NatPh..14..480G}.

\subsection{Energy Flux}
\begin{table*}
    \centering
        \caption{The computed parameters, photospheric cutoff frequency $(f_c)$, plasma $\beta$, absolute umbral average magnetic field ($\left |B_{umbra}\right |$), HMI Continuum density ($\rho _{cont}$), the corresponding energy flux ($F_{cont}$), and the Doppler velocity amplitude $(\left<v\right>_{dopp})$ obtained from the HMI Dopplergram data, are listed for all 20 active regions analysed along with their standard errors. The start time and NOAA number of each dataset are also listed. An asterisk symbol next to the NOAA number represents the cases where a light bridge is present within the sunspot.}
    \begin{tabular}{cccccccc} \hline 
    
        Observation          &NOAA &$f_c$& $\left|B_{umbra}\right|$& $\rho _{cont}$$\times 10^{-7}$& $\textrm{Plasma-$\beta$}$& $\left<v\right>_{dopp}$&$F_{cont}$\\
         Date \& Time&Number&($\textrm{mHz}$) &(G) &$(\textrm{$g\ cm^{-3}$})$&  &($\textrm{$m\ s^{-1}$}$)  &($\textrm{$kW\ m^{-2}$})$\\  \hline 
            
          2012 Aug 5 12:00 &11535&4.46 $\pm$  0.17&1534 $\pm$ 217&8.67 $\pm$ 2.45&2.62 $\pm$ 0.14&34.61 $\pm$ 11.57&3.57 $\pm$  2.59\\  
         2012 Sep 23  12:00&11575*&4.20 $\pm$ 0.36&1295 $\pm$ 192&5.53 $\pm$ 1.64&2.56 $\pm$ 0.13&35.88 $\pm$ 9.07&1.33 $\pm$ 0.78\\   
         2012 Oct 12 13:00&11586&4.44 $\pm$ 0.19&1379 $\pm$ 209&6.57 $\pm$ 1.99&2.63 $\pm$ 0.12&34.22 $\pm$ 6.92&3.90 $\pm$ 1.98\\   
        2012 Oct 19 12:00&11591& 4.82 $\pm$ 0.14&1525 $\pm$ 192&8.51 $\pm$ 2.14&2.60 $\pm$ 0.17&37.43 $\pm$ 13.67&5.34 $\pm$ 4.13\\  
        2012 Dec 4  12:00&11623*&4.19 $\pm$0.14&1474 $\pm$ 200&7.17 $\pm$ 1.95&	2.53 $\pm$ 0.11&39.47 $\pm$ 9.84&2.56 $\pm$ 1.45\\   
        2013 Feb 5 12:00&11665&4.47 $\pm$ 0.30&1447 $\pm$ 156&7.44 $\pm$ 1.60&	2.63 $\pm$ 0.13&33.81 $\pm$ 6.10&2.77 $\pm$ 1.17\\   
        2013 Sep 3  12:00&11836*&4.07 $\pm$ 0.35&1400 $\pm$ 204&6.57 $\pm$ 1.91&2.42 $\pm$ 0.12&36.67 $\pm$ 7.58&2.48 $\pm$ 1.12\\   
        2013 Nov 14 13:00&11895&3.94 $\pm$ 0.38&1339 $\pm$ 202&6.14 $\pm$ 1.85&	2.62 $\pm$ 0.13&35.71 $\pm$ 8.22&2.68 $\pm$ 1.48\\   
        2013 Dec 28 13:30&11934&4.67 $\pm$ 0.30&1282 $\pm$ 178&5.80 $\pm$ 1.61&	2.64 $\pm$ 0.16&35.30 $\pm$ 6.77&1.71 $\pm$ 0.81\\ 
       2014 Mar 18 12:00&12005&4.32 $\pm$ 0.15&1529 $\pm$ 232&8.81 $\pm$ 2.68&	2.65 $\pm$ 0.12&33.94 $\pm$ 8.70&3.22 $\pm$ 1.92\\   
        2014 Apr 13 12:00 &12032&4.25 $\pm$ 0.28&1497 $\pm$ 237&8.18 $\pm$ 2.59&2.54 $\pm$ 0.14&32.11 $\pm$ 6.66&3.44 $\pm$ 1.80\\   
        2014 May 3 12:00&12049*&4.59 $\pm$ 0.20&1571 $\pm$ 176&8.33 $\pm$ 1.87&	2.59 $\pm$ 0.09&45.46 $\pm$ 13.52&5.93 $\pm$ 3.77\\  
        2014 Jun 16 03:50&12090&3.97 $\pm$ 0.47&1374 $\pm$ 251&6.57 $\pm$ 2.40&	2.20 $\pm$ 0.13&36.70 $\pm$ 5.30&1.40 $\pm$ 0.65\\   
        2014 July 7  12:00&12109*&4.57 $\pm$ 0.24&1572 $\pm$ 310&9.12 $\pm$ 3.60&	2.61 $\pm$ 0.18&36.23 $\pm$ 16.66&3.95 $\pm$ 3.93\\  
        2014 July 28  12:00&12121*&4.51$\pm$0.21&1457 $\pm$ 181&6.96 $\pm$ 1.73&2.58 $\pm$ 0.09&43.15 $\pm$ 12.39&4.35 $\pm$ 2.73\\ 
        2014 Aug 11 12:00& 12135&4.58 $\pm$ 0.34&1441 $\pm$ 209&7.02 $\pm$ 2.04&	2.59 $\pm$ 0.12&33.84 $\pm$ 6.13&3.45 $\pm$ 1.60\\ 
        2014 Oct 14 12:00 &12186&4.33 $\pm$ 0.30&1282 $\pm$ 196&5.73 $\pm$ 1.75&	2.66 $\pm$ 0.13&31.50 $\pm$ 5.59&2.48 $\pm$ 1.16\\  
        2014 Dec 17 12:00 &12236*&4.30 $\pm$ 0.22&1285 $\pm$ 190&5.73 $\pm$ 1.70&2.59 $\pm$ 0.09&37.30 $\pm$ 6.91&1.77 $\pm$ 0.84\\   
        2015 Jan 29 10:30&12268*&4.91 $\pm$ 0.15&1535 $\pm$ 173&8.07 $\pm$ 1.82&	2.45 $\pm$ 0.14&34.70 $\pm$ 6.71&3.41 $\pm$ 1.53\\  
        2016 Jun 16 11:00& 12553 &4.40 $\pm$ 0.19&1774 $\pm$ 229&11.37 $\pm$ 2.89&2.59 $\pm$ 0.09&40.26 $\pm$ 9.74&6.66 $\pm$ 3.63\\ \hline
       
    \end{tabular}
    \label{tab:density}
\end{table*}{}

In general terms, the expression for wave energy flux can be written as \citep[e.g.,][]{2016ApJ...831...24K}:
\begin{equation}
    \vec{F} = \rho \left<v\right>^2 \vec{V_g}   +  (\vec{\left<v\right>} \times \vec{B}) \times \vec{\left<B\right>}.
    \label{flux}
\end{equation}
Here, $\rho$, $\left<v\right>$, $\vec{V_g}$, and $\vec{B}$ are the mass density, velocity amplitude, group velocity, and magnetic field strength, respectively. The first term on the right-hand side corresponds to the thermal kinetic energy flux while the second term represents the Poynting flux. Under low plasma $\beta$ conditions like that in the umbral atmospheres we are dealing with here, SMAWs propagate primarily along the magnetic field at local sound speed. Additionally, because SMAWs are longitudinal in nature, the velocity perturbations will be parallel to the magnetic field resulting in zero contribution to the Poynting flux. Therefore, Eq. \ref{flux} will become
\begin{equation}
    F = \rho  \left<v\right>^2 {c_s}.
    \label{energy}
\end{equation}
Utilising the height-dependent velocity amplitude, sound speed, and the density values obtained so far, we estimate the energy flux of SMAWs as a function of height following Eq. \ref{energy}. 
\begin{figure*}
    \centering
    \includegraphics[width = \linewidth]{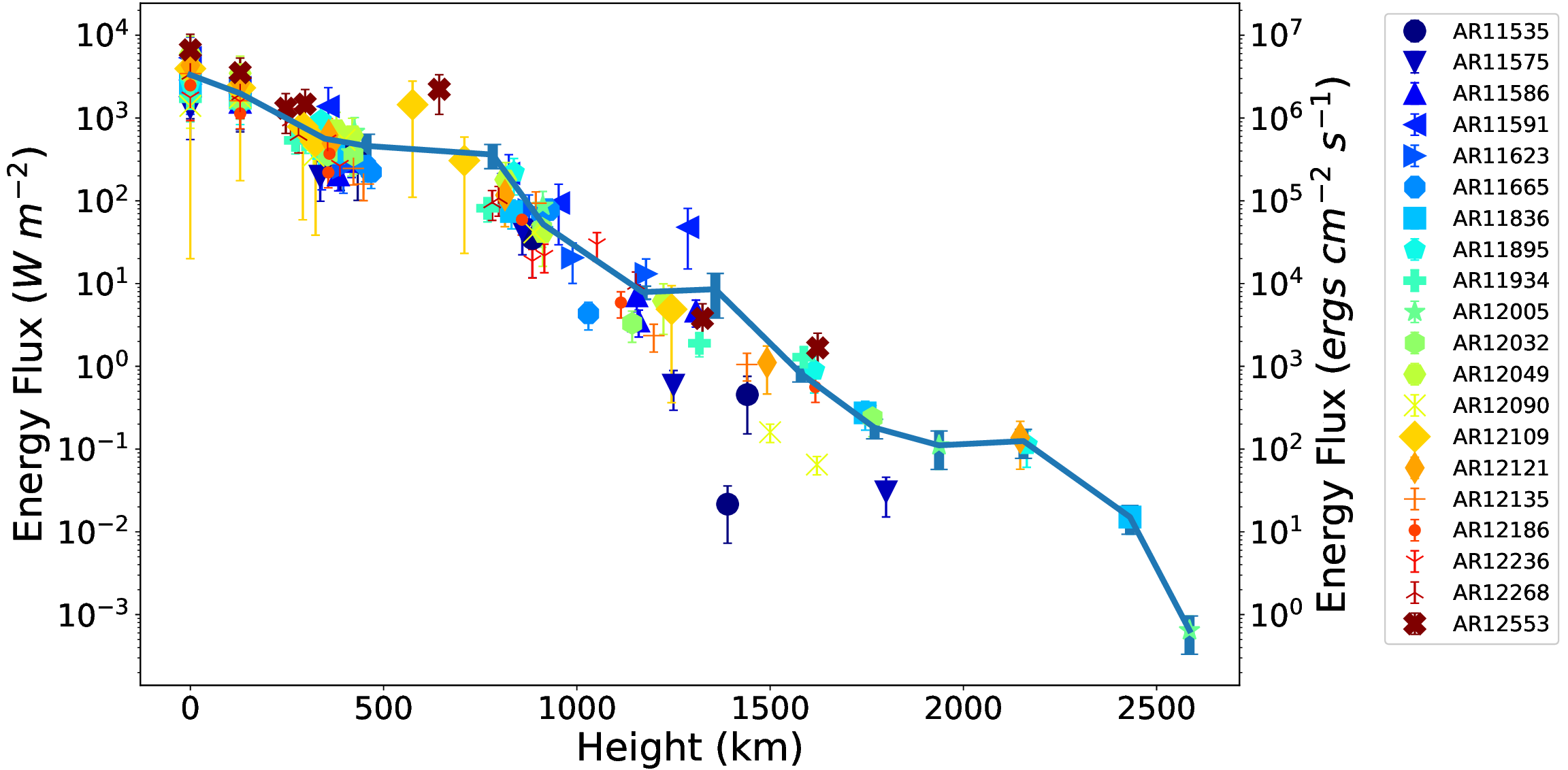}
    \caption{The height-dependent variation in the energy flux of 3-minute SMAWs starting from the HMI continuum height at 21 km in the photosphere. Different colors/symbols represent data from different active regions as listed. The solid line connects the average energy flux at an average formation height of all the data points within each 210 km block. The corresponding standard errors are shown as error bars.}
    \label{fig:flux}
\end{figure*}
The obtained values for all 20 active regions are shown in Fig. \ref{fig:flux}. Different colors/symbols represent data from different active regions. The vertical bar over each symbol denotes the error propagated from the uncertainty in corresponding velocity amplitude and density. A steady decrease in energy flux is readily visible from this plot. In order to find an average trend, similar to that in Fig. \ref{fig:ampheight}, within each block of 210 km, we compute the average energy flux and average formation height and connect them as a solid line in the figure. The respective uncertainties denote the standard errors.

\section{Discussion and Conclusions}\label{sec:conclusion}
We study the 3-minute propagating SMAWs in the umbral atmospheres of 20 different active regions using the multi-wavelength imaging data from SDO. The primary goal is to extract and understand the variation in wave amplitude and energy flux with height. We estimate amplitudes from the normalised Fourier power spectra and combine them with relevant densities and propagation speeds to derive the corresponding energy fluxes.

The average Fourier power spectra for each atmospheric channel highlight the dominant oscillatory modes at various altitudes, demonstrating the shift from 5 minute oscillations in the photosphere to 3 minute oscillations at higher altitudes (see Fig. \ref{fig:noise}). 
The discrepancy between the dominant oscillation frequencies at different atmospheric heights had sparked two different theories \citep[e.g., see review][]{2015LRSP...12....6K}. One theory suggests that the 3 minute oscillations in the upper atmosphere result from the formation of chromospheric or photospheric resonance cavities \citep[e.g.,][]{1982SoPh...79...19T,1983SoPh...82..369Z,1984MNRAS.207..731Z,2020NatAs...4..220J,2024MNRAS.529..967S}. Alternatively, the broader accepted theory says that the 3 minute oscillations are inherently present in the photosphere but have lower power levels and as they propagate upward they gradually dominate, as the 5 minute oscillations are filtered out by the atmospheric cutoff frequency \citep[e.g.,][]{2006ApJ...640.1153C,2017ApJ...836...18C}. Our power spectra indeed show a shift in peak frequency but cannot conclusively support either theory. 

The height-dependent variation in the amplitude of SMAWs reveals a complex pattern (see Fig. \ref{fig:ampheight}). The amplitudes initially increase up to a height of about 800 $\textrm{km}$ (rises from $\approx$ 24 to 120 $\textrm{$m\ s^{-1}$}$) before gradually decreasing at higher altitudes up to about 1800 km. This part of the trend is consistent with the previous results \citep[e.g.,][]{2017ApJ...847....5K,2024MNRAS.533.1166R}. The explanation is that the initial rise in amplitude is due to the fall in density, a consequence of energy conservation, and the subsequent decrease at higher altitudes is because of strong dissipation \citep{2010ApJ...719..357F}. However, interestingly, a secondary rise and decrease in wave amplitude is seen at altitudes above 1800 km. We conjecture this rise could be again due to a fall in density near the transition region and the subsequent decrease may be attributed to a strong damping via thermal conduction in the solar corona. We also note that the data at this altitude region is sparse so additional observations would be required to confirm this behavior.

Using the phase difference spectra, we estimate the cutoff frequency at HMI continuum height lies between 4 $\--$ 5 mHz. Combining this with the local sound speed and magnetic field strength, we estimate the density and, thereby, the energy flux. The energy flux shows a steady and monotonous decrease with height, consistent with previous findings \citep{2010ApJ...722..131F,2017ApJ...847....5K} although we show the trend here up to larger altitudes. The energy flux, from a value of $\approx$ 3.32 $\pm$ 0.50 $\textrm{$kW\ m^{-2}$}$ at 21 $\textrm{km}$, drops to $\approx$ 0.36 $\pm$ 0.12 $\textrm{$kW\ m^{-2}$}$ at $\approx$ 782 $\textrm{km}$ and further declines to $\approx$ 0.82 $\pm $ 0.18  $\textrm{$W\ m^{-2}$}$  near 1,580 $\textrm{km}$. At a height of 2,585 $\textrm{km}$, the energy flux is a mere (6.47 $\pm$ 3.16)$\times 10^{-4}$ $ \textrm{$W\ m^{-2}$}$, highlighting a substantial damping. These values are more or less in agreement with previous works, such as \citet{2017ApJ...836...18C}, which reported an average energy flux of 0.61 $\textrm{$kW\ m^{-2}$}$ at a height of 684 $\textrm{km}$, and \citet{2011ApJ...735...65F} who used MHD simulations to find the average acoustic energy flux at 725 $\textrm{km}$ for $\textrm{He\ I}$ line as 0.30 $\textrm{$kW\ m^{-2}$}$. As mentioned, this energy loss could be because of different mechanisms at different heights. The radiative damping and shock dissipation may be more effective at chromospheric heights while thermal conduction and viscosity may play a dominant role in the corona. 

We note that even after considering the highest energy flux in our sample (6.66 $\pm$ 3.63 $\textrm{$kW\ m^{-2}$}$), obtained for AR 12553, the energy would be inadequate to account for total chromospheric losses \citep[20 $\textrm{$kW\ m^{-2}$}$][]{1977ARA&A..15..363W}. The average Doppler velocity amplitudes we obtained from the HMI Dopplergram data are of the order of 30 $\--$ 45 $\textrm{$m\ s^{-1}$}$. Similar values were reported by \citet{1985ApJ...294..682L} and \citet{2017ApJ...836...18C} but these are much lower than those reported by some other studies \citep[e.g.,][]{1982ApJ...253..386L,2012A&A...539L...4S}, which found amplitudes greater than 100 $\textrm{$m\ s^{-1}$}$. Since the energy flux is dependent on the square of velocity amplitudes, if the amplitude is 100 $\textrm{$m\ s^{-1}$}$ or greater, the energy flux can be sufficient to maintain the chromosphere at such a high temperature. Moreover, it remains challenging to ascertain whether all the missing wave energy directly contributes to the heating of the local plasma. A substantial portion of wave energy may instead be transformed into other modes (fast or Alfv\'en) through processes such as mode conversion \citep[e.g.,][]{1992ApJ...391L.109S,2006MNRAS.372..551S}, which could result in dissipationless damping. Therefore, understanding these interactions is crucial for assessing the energy dynamics within the solar atmosphere.

\begin{acknowledgements}
The data used here are courtesy of NASA/SDO and the HMI and AIA science teams. S.K.P. is grateful to SERB for a startup research grant (No. SRG/2023/002623).
\end{acknowledgements}

\bibliography{flux}
\end{document}